# Direct Reconstruction of Distorted Signals and Images Using Shifts Methods

A.V. Novikov-Borodin *

*Abstract*— Mathematical methods of step-by-step and combined shifts are proposed for experimental data processing to reconstruct the measuring system impulse response distorted by shift-invariant blur. Proposed methods base on direct non-blind deconvolution without using approximations and integral transforms. Methods are fast and effective for accurate data reconstruction, which gives a possibility of increasing the effective resolution of measuring systems by mathematical methods up to physical limits without solving the expensive and quite difficult scientific and technical problems. Step-by-step and combined shifts methods supplement each other in data reconstruction at different distortions of signals, noise levels and data volumes. Methods may be adapted for reconstruction of multi-dimensional data. There are considered the restorations of 2D images blurred by uniform motion and distorted by functions, which may be factored, such as Gaussian-like functions. The comparative analysis of step-by-step and combined shifts methods is presented. Reconstruction inaccuracies are estimated. Examples of signal reconstructions and image restorations at different distortions are considered.

*Index Terms*— experimental data processing, direct non-blind deconvolution, signal reconstruction, image restoration

## I. INTRODUCTION

One of the basic problems in scientific investigations is to get the correct information about physical objects and processes of surrounding world. However, any experimental data are always distorted during detection, data treatments and transforms. Distortions are introduced not only due to the imperfection of measuring systems, but also due to the process of measurements itself. Indeed, the fact, that any measurement cannot be done in a moment, but takes some time, already pre-defines the overlapping the hypothetical `ideal' momentary signals from investigated objects during the time of measurements. Thus, such overlaps initially exist in experimental data for any practical measurements. It is especially evident in statistical measurements, because they need a long period of time to collect the information.

When overlaps during measurements are described by the function $S(x)$, the response $H(x)$ of the measuring system is a convolution of its impulse response $h(x)$ and the superposition function $S(x)$:

$$H(x) = S * h = \int S(\xi) h(x-\xi) d\xi. \qquad (1)$$

In fact, the equation (1) is an integral transform with the kernel $S(x)$. Taking into account the distortion character of the superposition function $S(x)$, one may call it a blur kernel.

Thus, to get the correct information from experimental data means to reconstruct the impulse response $h(x)$ from distorted data $H(x)$ using the operation reversed to convolution (1), i.e. using the deconvolution. Deconvolution is usually called blind, if the blur kernel $S(x)$ is unknown, and non-blind otherwise. Even non-blind deconvolution is considered in mathematics as the ill-posed problem [1], because in general case, there are no enough information in equation (1) for reconstruction. The existence of noise in experimental data worsens the situation, so, in practice, even non-blind deconvolution problem cannot be solved accurately, and this problem remains an active and challenging research topic [2,3].

Although, generally, the accurate deconvolution is impossible, there exists a possibility of data reconstruction in many particular cases important in practice. For example, as it was shown in [4], one may reconstruct the statistical experimental data of neutron spectrometers using the step-by-step shifts method. The possibility of accurate reconstruction depends on many reasons: the kind of the blur kernel and an accuracy of its definition, the capacity of experimental data, noise level in them, etc. That is why it needs the development of different methods effective in one or another case.

The possibility of reconstruction using the step-by-step shifts method is analyzed in this paper and there is proposed the combined shifts method, which gives a possibility to overcome some principal limitations of the step-by-step method. Proposed step-by-step and combined shifts methods supplement each other in data reconstruction with different superposition functions, noise levels and data volumes. A possibility of accurate reconstruction is important in experimental data processing, because in fact it means the increasing the resolution of measuring systems. There is considered a possibility of renewal the images blurred by uniform motion and distorted with Gaussian-like kernel using proposed methods.

* A.V. Novikov-Borodin is with the Experimental Physics Department, Institute for Nuclear Research (INR), Russian Academy of Sciences, Moscow, 117312 Russia (e-mail: novikov.borodin@gmail.com).

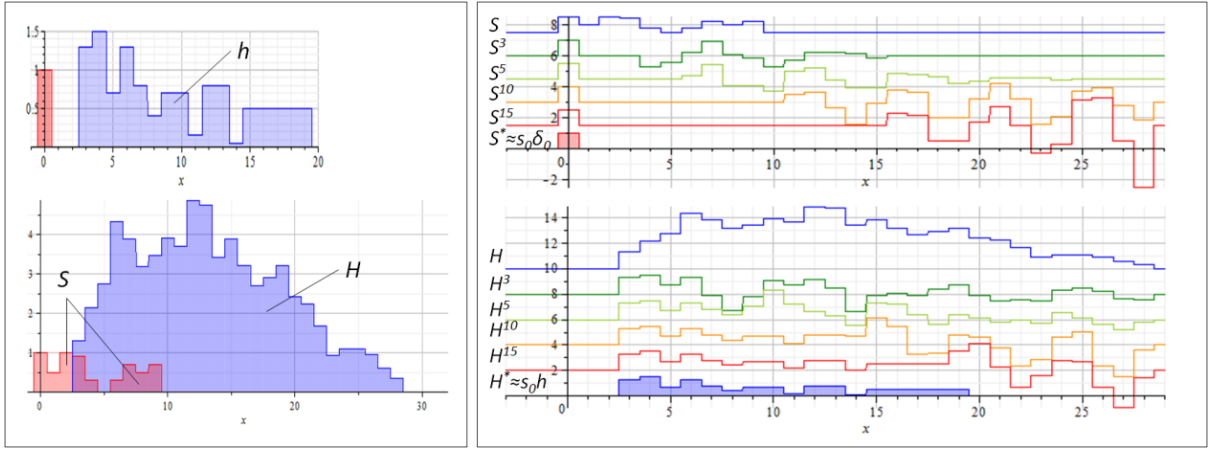

**Figure 1.** Reconstruction of the impulse response $h$ from the given system response $H$ with the blur kernel $S$ using the step-by-step shifts method.

## II. METHOD OF STEP-BY-STEP SHIFTS

On discrete intervals introduced on $x$-axis the functions $S(x)$ and $h(x)$ from (1) may be represented as: $S = \sum_k s_k \delta_k$ and $h = \sum_m h_m \delta_m$ respectively, where $\delta_i$ is the Kronecker symbol differing from 0 and equal to 1 only on the $i$-th interval. In this case the system response $H$ will be:

$$H = \sum_p H_p \delta_p = \sum_k \sum_m s_k h_m \delta_{k+m}. \quad (2)$$

Equations (1) and (2) are linear in relation to the function $S$. Indeed, if $S(x) = \alpha_1 S_1(x-x_1) + \alpha_2 S_2(x-x_2) + \cdots$, so $H(x) = \alpha_1 H_1(x-x_1) + \alpha_2 H_2(x-x_2) + \cdots$. Step-by-step shifts method, introduced in [4], uses this linearity for direct reconstruction of data by means of iteration equations:

$$S^{n+1} = S^n - a_n S^n_{(n+1)}, \; H^{n+1} = H^n - a_n H^n_{(n+1)}. \quad (3)$$

Here $S^n = \sum_k s^n_k \delta_k$ and $H^n = \sum_k \sum_m s^n_k h^n_m \delta_{k+m}$ are respectively the superposition function and the system response on the $n$-th iteration step, $S^n_{(l)} = \sum_k s^n_k \delta_{k+l}$ and $H^n_{(l)} = \sum_k \sum_m s^n_k h^n_m \delta_{k+m+l}$ are the same functions shifted to the right on $l$ discrete intervals, $S^0 = S$, $H^0 = H$, $a_n = s^n_{n+1}/s_0$.

According to algorithm (3), the first $n$ coefficients in the superposition function $S^n$ on $n$-th iteration step is equal to zero: $s^n_1 = s^n_2 = \cdots = s^n_n = 0$, so one has:

$$\begin{cases} S^n = s_0 \delta_0 + \sum_k s^n_{k+n+1} \delta_{k+n+1} \\ H^n = s_0 \sum_m h_m \delta_m + \sum_k \sum_m s^n_{k+n+1} h^n_m \delta_{m+k+n+1} \end{cases}. \quad (4)$$

Thus, in the interval $[0,n]$: $S^n = s_0 \delta_0$ and $H^n = s_0 \sum_m h_m \delta_m = s_0 h$. It means that on this interval the impulse response of the measuring system is reconstructed.

An example of reconstruction of the impulse response $h$ from the given system response $H$ with the blur kernel $S$ by using the step-by-step shifts method (3)-(4) is presented on Figure 1. There are shown on the diagrams the superposition functions and responses on initial, 3-d, 5-th, 10-th, 15-th and last iteration steps. The reconstruction error $err = H^* - s_0 h$ on the interval $[0,20]$ did not exceed $2.5 \cdot 10^{-15}$.

Using the step-by-step shifts method, one can reconstruct as many elements of the impulse response as needed. It gives a possibility to process the experimental data of unlimited length. One can see on diagrams of the Figure 1 that the first $n$ elements of the impulse response on the $n$-th iteration step are being successively reconstructed.

Let's consider two impulse responses overlapped with the blur kernel $S = s_0 \delta_0 + s_l \delta_l$. If on some iteration steps (3) $a_n = s^n_{n+1}/s_0 = 0$, functions $S$ and $H$ do not change, so the iteration equations (3) may be modified as:

$$S^{n+1} = S^n + a^{2^n} S^n_{(2^n l)}, \; H^{n+1} = H^n + a^{2^n} H^n_{(2^n l)}, \quad (5)$$

where $a = s_l/s_0$, $n = 1, 2, \ldots$, $S^0 = S$, $H^0 = H$, $S^1 = S^0 - a S^0_{(l)}$, $H^1 = H^0 - a H^0_{(l)}$.

The modified method reduces considerably the number of iteration steps needed for reconstruction. If $N$ is a number of iteration steps needed for reconstruction using the step-by-step shifts method, so a number $N^*$ of iteration steps needed for the same reconstruction using the modified method will be: $N^* = \log_2(N/l)$.

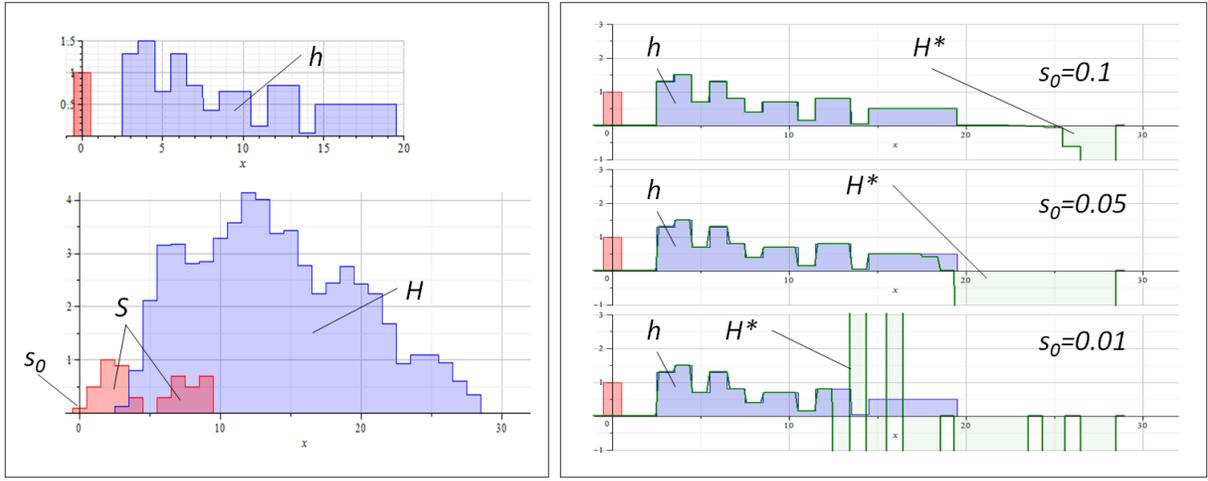

**Figure 2.** The error increasing with decreasing the coefficient $s_0$ in the function $S$ during the step-by-step reconstruction.

One may try to use the modified method for modeling the response of the measuring system on the initializing pulse of less duration. In fact, it means a possibility of increasing the effective measurement resolution up to physical limits. An effort to get the resolution higher physical limits leads to iteration sequence divergence.

Theoretically, the method of step-by-step shifts gives a possibility of reconstruction of the impulse response with any blur kernel $S = \sum s_k \delta_k$, but in practice, if some of the coefficients $s_k > s_0$, the coefficients $a_n$ in (3) or $a^{2^n}$ in (5) will increase quickly. It leads to extremely high growth of the numerical calculation errors and to the divergence of the iteration sequence. It is shown on Figure 2 that the reconstruction errors increase with decreasing the coefficient $s_0$ of the blur kernel on Figure 1 from 0.1 to 0.01.

Indeed, if $S = s_0 \delta_0 + s_l \delta_l$, where $s_l / s_0 = a > 1$, so even if $s_l$ is greater than $s_0$ only on one percent, i.e. if $a = 1.01$, the coefficient $a^{2^n}$ just on the twentieth iteration step will be $a^{2^{20}} \approx 1.01^{10^6} \approx 1.94 \cdot 10^{4531}$. Thus, if $N_{max}$ is the maximal permissible value for digital calculations, one can reconstruct only first $M_{max}$ elements of the impulse response:

$$M_{max} \approx l \frac{\ln(N_{max})}{\ln(s_{max}/s_0)}. \qquad (6)$$

The formula (6) may be used for estimations for any blur kernel. For example, one can see on Figure 2 that $s_{max} = s_2 = 1$, and $M_{max} \approx 13, 19-20, 26$ when $s_0 = 0.01, 0.05, 0.1$ respectively. It is in a good agreement with the formula (6), which gives: $M_{max} \approx 13, 19.98, 26$.

In case of two impulses $S = s_0 \delta_0 + s_l \delta_l$ with $s_l / s_0 > 1$, one can make the reconstruction in relation to term $s_l \delta_l$, shifting the iteration sequence not to the right, but to the left. In this case $s_0 / s_l = a < 1$ and a coefficient $a^{2^n}$ in iteration equation (5) will decrease quickly. Indeed, if $a = 0.99$, just on 9-th iteration step the reconstruction error is 0.6 %, and on 10-th it is $3.4 \cdot 10^{-3}$ %.

However, if the maximal term of the blur kernel is not the first or the last one, one cannot simultaneously zeroing the coefficients on the left and right from it using the step-by-step shifts method. Moreover, the step-by-step shifts method is very sensitive to noise ($H_{noise}$) in experimental data ($H$). It is shown on Figure 3 that even if the noise level is 1-5% the iteration sequence is divergent and the reconstruction is impossible.

These drawbacks are very critical for statistical measurements, because the reconstruction in relation to non-maximal element means the reconstruction of measurements with small statistics. The method being proposed to eliminate such problems bases on the same method of step-by-step shifts, but uses a combination of a few shifts.

### III. METHOD OF COMBINED SHIFTS

Let's consider the blur kernel $S^*$, which is a linear combination with coefficients $\mu_i$ of shifted functions $S = s_0 \delta_0 + s_1 \delta_1 + s_2 \delta_2$, where $s_{max} = s_1$:

$$S^* = \mu_{-1} S_{(-1)} + \mu_0 S_{(0)} + \mu_1 S_{(1)}. \qquad (7)$$

Here $S_{(i)} = s_0 \delta_i + s_1 \delta_{i+1} + s_2 \delta_{i+2}$ is the function $S$ shifted on $i$ discrete intervals, and coefficients $\mu_i$ are chosen

A.V. Novikov-Borodin

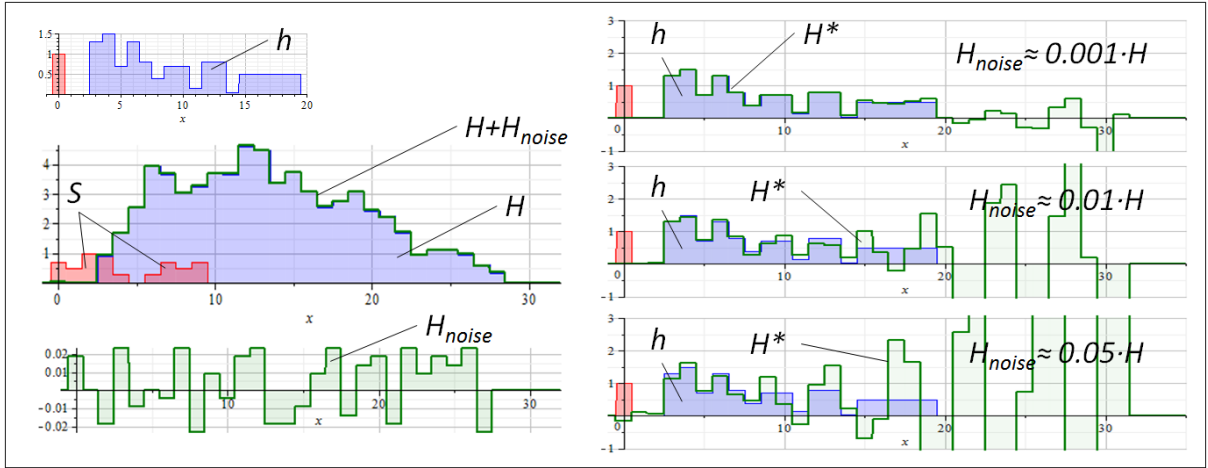

**Figure 3.** The reconstruction error of the step-by-step method with noise levels from 0.1 to 5 %.

to get the coefficients $s_0^*, s_2^* = 0$ and $s_1^* = 1$ in the function $S^*$. Representing $S^* = s_{-1}^* \delta_{-1} + \delta_1 + s_3^* \delta_3$ as a vector $\mathbf{s} = \begin{bmatrix} s_{-1}^* & 0 & 1 & 0 & s_3^* \end{bmatrix}$ and $\boldsymbol{\mu} = \begin{bmatrix} \mu_{-1} & \mu_0 & \mu_1 \end{bmatrix}$, one can rewrite the equation (7) in the matrix form:

$$\mathbf{s} = \boldsymbol{\mu} \cdot \begin{bmatrix} s_0 & s_1 & s_2 & 0 & 0 \\ 0 & s_0 & s_1 & s_2 & 0 \\ 0 & 0 & s_0 & s_1 & s_2 \end{bmatrix}. \quad (8)$$

As far as in this example the values $s_{-1}^*, s_3^*$ are not important, to find the unknown coefficients $\mu_i$ one can solve the reduced matrix equation:

$$\mathbf{e} = \begin{bmatrix} 0 & 1 & 0 \end{bmatrix} = \boldsymbol{\mu} \cdot \begin{bmatrix} s_1 & s_2 & 0 \\ s_0 & s_1 & s_2 \\ 0 & s_0 & s_1 \end{bmatrix} = \boldsymbol{\mu} \cdot \boldsymbol{\Sigma}. \quad (9)$$

According to applied conditions, the determinant of the quadratic shifts matrix $\boldsymbol{\Sigma}$ is not equal to zero: $\det(\boldsymbol{\Sigma}) \neq 0$, so the inverse matrix $\boldsymbol{\Sigma}^{-1}$ exists, and the equation (9) always has a solution:

$$\boldsymbol{\mu} = \mathbf{e} \cdot \boldsymbol{\Sigma}^{-1}. \quad (10)$$

Thus, using a linear combination (7) of shifted functions $S$ with coefficients (10) one can get the new kernel $S^*$, with which the coefficients on the right and left from the chosen term equal to zero.

Analogously, using the linear combination of functions $S$ shifted on the right and left on $L$ intervals with coefficients $[\mu_{-L}, \ldots, \mu_0, \ldots, \mu_L]$, one can get the kernel $S^* = \mu_{-L} S_{-L} + \cdots \mu_0 S_0 + \cdots \mu_L S_L$, with which $L$ coefficients on the right and left from the chosen term $s_C$ equal to zero.

In this case, if $L > M$, where $M$ is a number of terms in impulse response $h = \sum_{m=0}^{M} h_m \delta_m$, on interval $[C, L+C]$ the response $H^*$ corresponding to $S^*$:

$$\begin{cases} H^* = \mu_{-L} H_{(-L)} + \cdots \mu_0 H_{(0)} + \cdots \mu_L H_{(L)} \\ \boldsymbol{\mu} = \mathbf{e} \cdot \boldsymbol{\Sigma}^{-1} \end{cases} \quad (11)$$

coincides with $h$: $H^* = h$. Exactly this algorithm of reconstruction is being called a method of *combined shifts*. Here, vector $\mathbf{e}$ with dimension $2L+1$ has only one nonzero element: $e_{L+1} = 1$, and the matrix $\boldsymbol{\Sigma}^{-1}$ is inverse to the quadratic $2L+1 \times 2L+1$ matrix $\boldsymbol{\Sigma} = \{\Sigma_{ij}\}$: $\Sigma_{ij} = s_{C+j-i}$, $i, j = 1, \ldots, 2L+1$.

One may find the shifts coefficients $\boldsymbol{\mu}$ without finding the inverse matrix, but using well-known methods of solving the system of linear equations $\boldsymbol{\mu} \cdot \boldsymbol{\Sigma} = \mathbf{e}$, where the coefficient at the chosen term needs to be maximal: $s_C = s_{\max}$ for better convergence.

The combined shifts method gives a possibility to reconstruct the impulse response with any blur kernel $S$, but dimensions $K$ and $M$ of functions $S$ and $h$ have to be finite. Also the system response $H$ has to be completely determined, i.e. its dimension needs to be more than $K + M$.

Using the combined shifts method one can easily reconstruct the impulse response in cases shown on Figure 2, where the step-by-step method was divergent. Examples of such reconstruction using the combined shifts method are presented on Figure 4. When $s_0 = 0.01$ the reconstruction error $err = H^* - h$ is $3.0 \cdot 10^{-12}$, and when $s_0 = 0.001$ it is less than $3.0 \cdot 10^{-8}$.

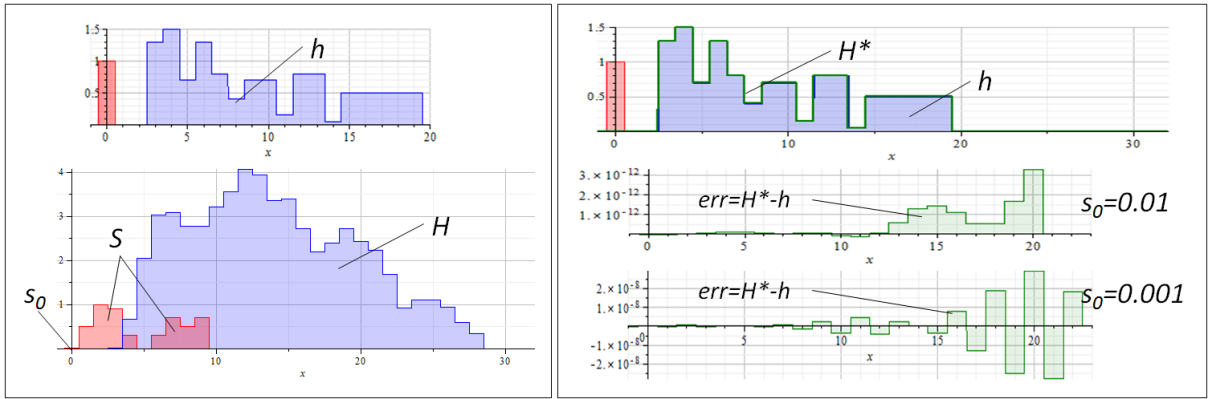

**Figure 4.** Combined shifts reconstruction of the impulse response $h$ with small first coefficient of the blur kernel $S$.

The combined shifts method can also solve the reconstruction problem when the blur kernel is Gaussian. Examples of such reconstruction using step-by-step (A) and combined (B) shifts methods are shown on Figure 5. The reconstruction error of combined shifts method (C) is less than $4.0 \cdot 10^{-12}$, while the iteration consequence of the step-by-step method is divergent. The possibility to choose the maximal term in the blur kernel for reconstruction is an important advantage of the combined shifts method in case of statistical measurements, because it means the reconstruction in relation to the term with maximal statistic. It determines the possibility of reconstruction with relatively high noise levels in experimental data.

It is shown on Figure 6 the results of reconstruction of the impulse response $h$ using step-by-step (A) and combined (B) shifts methods with noise level ($H_{noise}$) about 1.0 % in the experimental data $H$. While the iteration sequence of the step-by-step method is divergent (A), the reconstruction error $err = H^* - h$ of the combined shifts method (B) is approximately the same as the initial noise level $H_{noise}$.

The method of combined shifts gives a possibility to model the experimental data $H$ with some optimal blur kernel. It is enough for this to specify the vector $\mathbf{e}$ in equation (11) in needed form. Sometimes when the reconstruction of the impulse response is impossible, it may be the only possibility. Thus, the combined shifts method makes possible to reconstruct the impulse response in relation to element with the maximal statistic, to model the system response with the optimal kernel or from the initializing signal of lower duration. The last, in fact, means the increasing the effective resolution of the measuring system.

However, one may use the combined shifts method only if the blur kernel and the impulse response are defined on the finite intervals. One also needs to take into account that the combined shifts method operates with high dimension matrices, so needs a lot of calculations, and it may lead to error increasing during reconstruction.

The general method of solution the reconstruction problem of convolved experimental data does not exist. The possibility of reconstruction depends on the kind of the blur kernel, on accuracy of its definition, on experimental data volume, noise level and many other reasons. It needs the elaboration the different methods effective in one or another case. Methods of step-by-step and combined shifts supplement each other in reconstruction at different conditions.

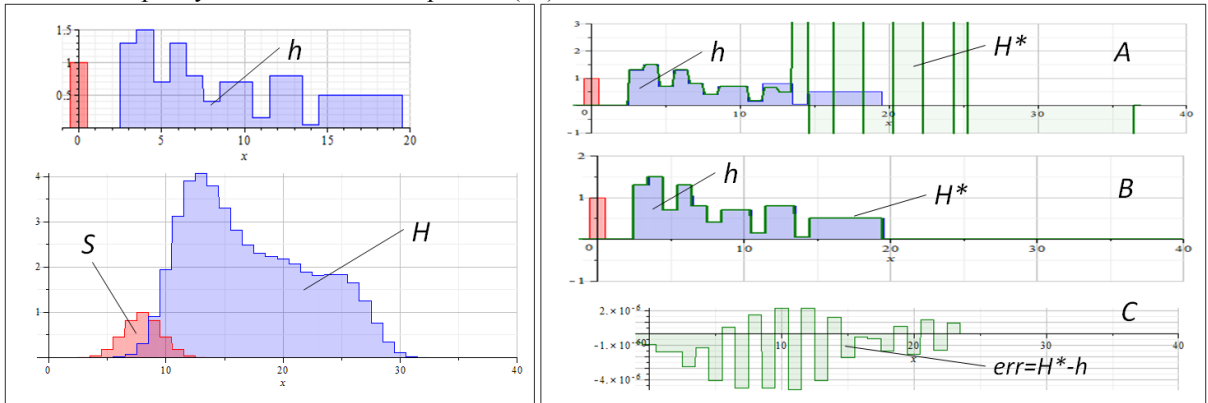

**Figure 5.** Reconstruction of the impulse response $h$ with Gaussian blur kernel $S$ using the step-by-step (A) and combined (B) shifts methods.

A.V. Novikov-Borodin

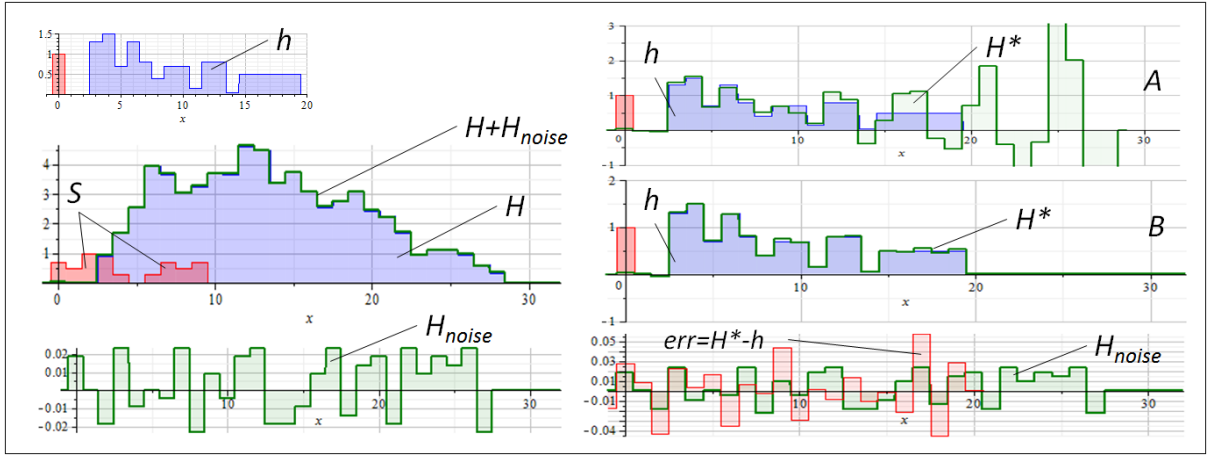

**Figure 6.** Reconstruction of the impulse response $h$ with noise $H_{noise}$ in experimental data $H$ using step-by-step (A) and combined (B) shifts method.

## IV. Image Restoration

In multi-dimensional case the result $H(\mathbf{x})$ of convolution of the image $I(\mathbf{x})$, $\mathbf{x} = (x_1, x_2, \ldots)$ and the blur kernel $S(\mathbf{x})$ may be represented as:

$$H(x_1, x_2, \ldots) = S * I = \\ = \iiint S(\xi_1, \xi_2, \ldots) I(x_1 - \xi_1, \ldots) d\xi_1 d\xi_2 \ldots \quad (12)$$

In discrete form the functions $S(\mathbf{x})$ and $I(\mathbf{x})$ are:

$$S = \sum_{k_1, k_2, \ldots} s_{k_1 k_2 \ldots} \delta_{k_1 k_2 \ldots} \quad \text{and} \quad I = \sum_{m_1, m_2, \ldots} I_{m_1 m_2 \ldots} \delta_{m_1 m_2 \ldots},$$

where $\delta_{i_1 i_2 \ldots}$ is a multi-dimensional Kronecker symbol differed from 0 and equal to 1 only in the volume bounded by $i_1, i_2, \ldots$-th discrete intervals on axes $x_1, x_2, \ldots$ respectively. In this case the response $H(\mathbf{x})$ is:

$$H = \sum_{k_1, k_2, \ldots; m_1, m_2, \ldots} s_{k_1 k_2 \ldots} I_{m_1 m_2 \ldots} \delta_{k_1 + m_1, k_2 + m_2, \ldots}. \quad (13)$$

The considered one-dimensional methods of step-by-step and combined shifts may be effectively used for reconstruction of multi-dimensional images whenever it is possible to reduce the equations (12) and (13) to one-dimensional case or to a combination of one-dimensional cases.

The simplest case of such reduction is a parallel shift, i.e. smearing the image $I(\mathbf{x})$ or objects in it along with one of the image axes. Let's consider without losing the generality, that the smearing is along the axis $x = x_1$. Equations (12) and (13) are just being transformed to one-dimensional equations (1)-(2):

$$H(x, x_2, \ldots) = \int S(\xi) I(x - \xi, x_2, \ldots) d\xi,$$

$$H = \sum_{p, p_2, \ldots} H_{p p_2 \ldots} \delta_{p p_2 \ldots} = \sum_{k, m} s_k I_{m m_2 \ldots} \delta_{k+m, m_2, \ldots}. \quad (14)$$

In this case the methods of step-by-step (3), (5) and combined (11) shifts may be used for image reconstruction. If images are color, each color layer needs to be reconstructed in the same way.

It is shown on Figure 7 the diagrams of successive reconstruction of two-dimensional color image $I(x, y)$ $79 \times 100$ pixels using the step-by-step shifts method on 0, 10, 20, 40 and 100-th iteration steps (3). The distorted image $H$ is smeared along with the $x$–axis with the blur kernel $S(x)$. The reconstruction error $err = H^* - I$ does not exceed $5 \cdot 10^{-15}$.

If the direction of the image shift does not coincide with the image axes, the simplest way of reconstruction is to rotate the image to match one of the axes with the shift direction. This way, however, one needs to take into account the increasing of discretization errors during axes rotation.

The important case of blurred images is images or objects in them smeared by uniform motion. It occurs, for example, during the registration of fast moving objects. Usually one needs only few iteration steps of the modified method (5) to restore such images. For example, to reconstruct the image on Figure 8 smeared by uniform motion with function $S = \delta_{0-19}$ one needs only 4 iteration steps using the modified method instead of 100 steps needed for restoration of the same image on Figure 7. Indeed, shifting the function $S$ on one interval and subtracting $S_{(1)}$ from $S$, one gets the kernel $S^1 = S - S_{(1)} = \delta_0 - \delta_{20}$ corresponding to two overlapped images. It is suitable for image reconstruction using the modified method (5). It is shown on Figure 8 that one needs only three additional iteration steps to put distortions out of the image.

Another important way to use considered one-dimensional shifts methods for image reconstruction is to reduce the multi-dimensional equations (12), (13) to a combination of one-dimensional ones (1), (2). It is possible if the function $S$ may be factored as: $S(\mathbf{x}) = S_1(x_1) S_2(x_2) \cdots S_n(x_n)$. In this case the equation (12) may be represented as:

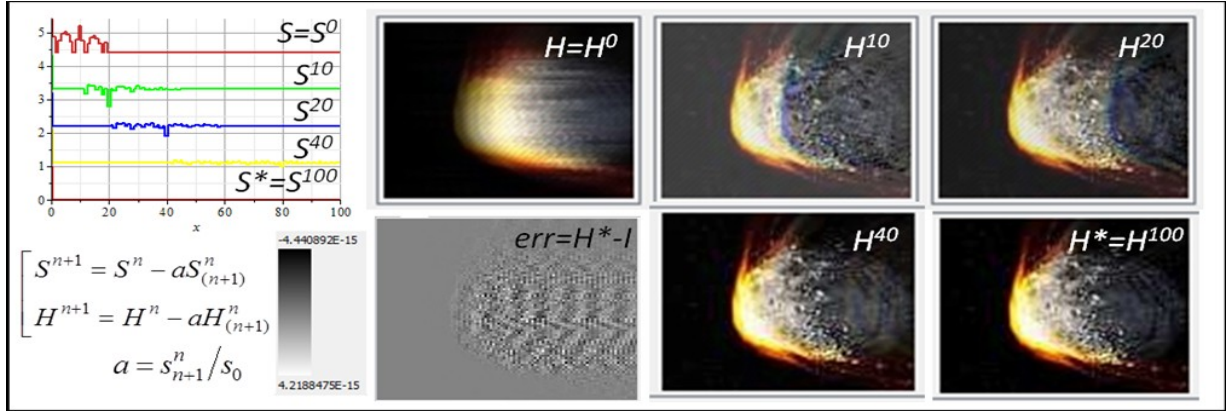

**Figure 7.** Restoration of 2D color image $I(x,y)$ from the image $H$ smeared with function $S(x)$ using the step-by-step shifts method.

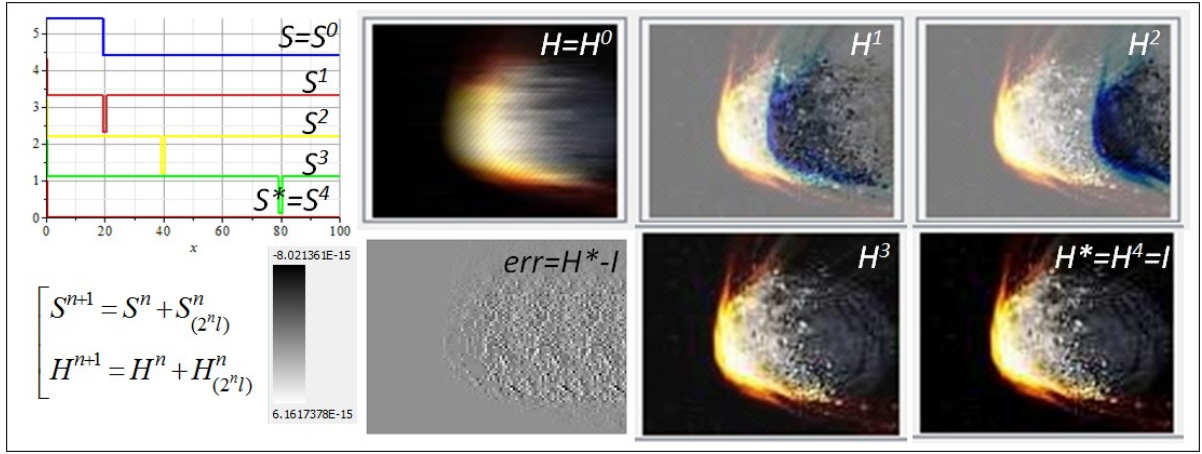

**Figure 8.** Restoration of 2D color image $I(x,y)$ from the image $H$ smeared with uniform motion using the modified method.

$$H(x_1, x_2, \ldots) =$$
$$= \int d\xi_1 S_1(x_1 - \xi_1) \int d\xi_2 S_2(x_2 - \xi_2) \cdots \quad (15)$$
$$\cdots \int d\xi_n S_n(x_n - \xi_n) I(\xi_1, \xi_2, \ldots, \xi_n)$$

One can rewrite equation (15) in the form:

$$\begin{bmatrix} H(x_1, \ldots, x_n) = \int d\xi_1 S(x_1 - \xi_1) I_1(\xi_1, x_2, \ldots); \\ I_1(x_1, \ldots, x_n) = \int d\xi_2 S(x_2 - \xi_2) I_2(x_1, \xi_2, \ldots); \\ \ldots \\ I_{n-1}(x_1, \ldots, x_n) = \int d\xi_n S(x_n - \xi_n) I(x_1, \ldots, \xi_n); \end{bmatrix} \quad (16)$$

and restore the image by consequent reconstruction of functions $I_1(\mathbf{x}), I_2(\mathbf{x}), \ldots, I_n(\mathbf{x}) = I(\mathbf{x})$ using one-dimensional shifts methods on corresponding coordinates.

An order of reconstructed functions may be random. For example, in two-dimensional case there are two ways of image restoration – through $I_x(x,y)$ or $I_y(x,y)$:

$$\begin{bmatrix} H(x,y) = \int d\xi S_x(x-\xi) I_x(\xi, y) \\ I_x(x,y) = \int d\eta S_y(y-\eta) I(x, \eta) \\ H(x,y) = \int d\eta S_y(y-\eta) I_y(x, \eta) \\ I_y(x,y) = \int d\xi S_x(x-\xi) I(\xi, y) \end{bmatrix} \quad (17)$$

It is shown on Figure 9 an example of restoration the image distorted by the two-dimensional Gaussian-like blur kernel $S(x,y) = \exp\left[-\frac{1}{2}\left(\frac{x^2}{\sigma_x^2} + \frac{y^2}{\sigma_y^2}\right)\right]$, which may be factored as: $S(x,y) = \exp\left(-\frac{x^2}{2\sigma_x^2}\right)\exp\left(-\frac{y^2}{2\sigma_y^2}\right)$. Arrows show two possible ways of restoration the image $H(x,y)$ through the reconstruction of images $I_x(x,y)$ or $I_y(x,y)$ blurred by $x$ or $y$ respectively. The reconstruction error $err = I^*(x,y) - I(x,y)$ did not exceed 1.5 %, while a difference between images reconstructed

A.V. Novikov-Borodin

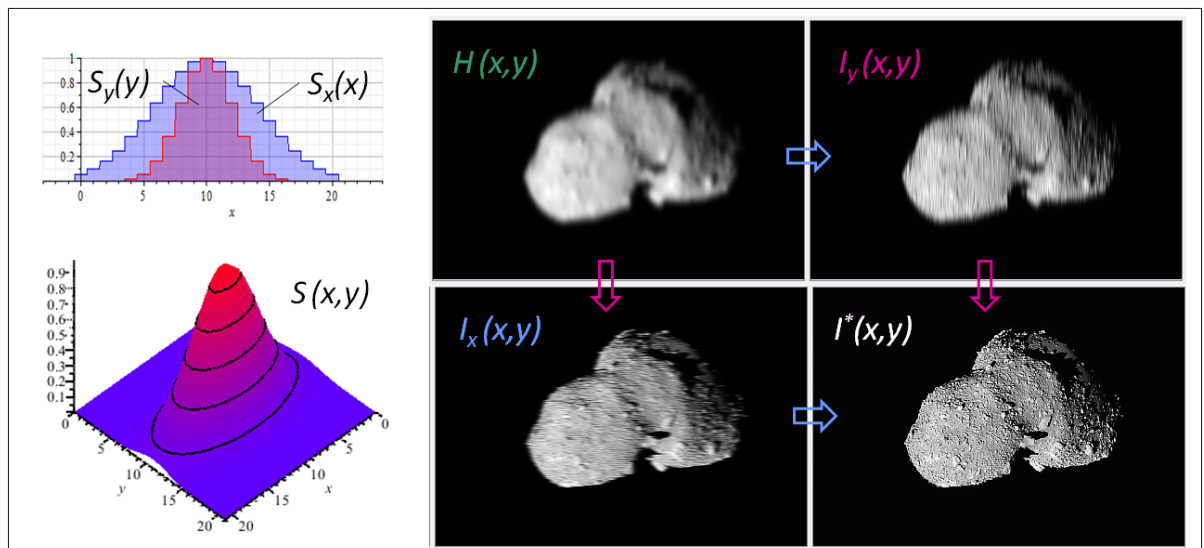

**Figure 9.** Restoration the image $I(x,y)$ from $H(x,y)$ blurred by 2D Gaussian function using combined shifts method.

by different ways $err = I^*_{xy}(x,y) - I^*_{yx}(x,y)$ was less than $2 \cdot 10^{-4}$ %.

The distortion with Gaussian kernel is very important in image restoration, because corresponds the random deviations of objects being registered or of the measuring equipment from initial fixed position.

## V. Conclusion

Proposed mathematical methods of step-by-step and combined shifts give a possibility to eliminate distortions in experimental data caused by different instabilities and to optimize the measurements. Optimization means the possibility of increasing the resolution of measurements right up to physical constrains, what is an alternative to expensive technical methods conjugated with hard-to-solve scientific and technical problems.

The general method of reconstruction the convolved data does not exist. The possibility of reconstruction depends on many reasons, such as the kind of superposition function and the accuracy of its definition, the volume of experimental data and the noise level in them and many others. It needs different methods for effective reconstruction in one or another particular case. Methods of step-by-step and combined shifts supplement each other at different conditions.

The important feature of these methods is a possibility of restoration the multi-dimensional data, in particular, images or objects in them smeared in parallel shifting or blurred with Gaussian-like functions.

Due to their simplicity and efficiency the proposed methods may be used in many different fields of experimental data processing, for reconstruction of digital signals and images, in development of deconvolution methods in mathematics.